\documentclass[12pt]{article}

\newcommand{\alg}[1]{\mathcal{#1}}
\newcommand{\val}[1]{\langle{#1}\rangle}
\newtheorem{theorem}{Theorem}

\begin{document}
\title{Beables in Algebraic Quantum Mechanics
\footnote{Forthcoming in H. Brown, J. Butterfield and C. Pagonis (eds.),
\emph{From Physics to Philosophy: Essays in Honour of Michael Redhead},
Cambridge University Press.}}
\author{Rob Clifton\footnote{Department of Philosophy, 1001 Cathedral of
Learning, University of
Pittsburgh, Pittsburgh, PA\ 15260 (email: rclifton+@pitt.edu).}}
\date{November 1997}
\maketitle

Michael Redhead is one of the foremost advocates of the
tenability of scientific realism in the domain of quantum theory.
Particularly inspiring is his deep physical
knowledge and intuition, combined with the uncanny ability he has to
tease out the essence of a conceptual problem in physics from amidst
the often bewildering and mathematically daunting literature.
It is a pleasure to be able to offer this piece in honour
of the (self-described) `Cantabridgian dinosaur's' retirement, and to
express my desire
that he not become extinct just yet!

\section{Beables versus `Observables'}

A good deal of Michael's work has focussed on
articulating the pitfalls of adopting a `simple realism of possessed
 values' in quantum mechanics, which is put under pressure by the
 no-go theorems of Kochen-Specker and Bell.  While I suspect Michael's views
 on the matter are still tentative and exploratory, in his
 recent book
 \emph{From Physics to Metaphysics} (1995a, Ch. 3) Michael
 appears to favour van
 Fraassen's (1973) idea of
  securing determinate values for all observables by `ontologically
contextualizing'
  physical magnitudes.  The idea
  is to let any given degenerate self-adjoint operator on a system's
 Hilbert space represent more than one magnitude
 of the system.
 Each magnitude is distinguished from the others by the functional relations
  its values have to different complete commuting sets of self-adjoint
  operators of which the given self-adjoint operator is a member.  Thus,
  to pick out a physical magnitude it is not enough to know that
  its statistics are represented by tracing
  the density operator of the system with some particular
  self-adjoint operator; for the degenerate operators, one must
  also
   pick a context of definition for
  the magnitude being measured, specified by some complete commuting set.

  Formally, this is enough to prevent Kochen-Specker
  contradictions.  But for Michael the real payoff is that
  it yields a novel holistic interpretation of quantum nonlocality
  (1995a, pp. 86-7).
  Take the example of two correlated spin-$1$ particles in the
  entangled singlet state
  \begin{equation}
  -3^{-\frac{1}{2}}[|S_{1x}=0\rangle|S_{2x}=0\rangle
  -|S_{1y}=0\rangle|S_{2y}=0\rangle
  +|S_{1z}=0\rangle|S_{2z}=0\rangle]
  \end{equation}
  for which
  Michael and a former student were able to supply the first
  purely algebraic proof of Bell's theorem (Heywood and Redhead
  1983).
  Since the self-adjoint operator $S_{1n}^{2}$ pertaining to any
  (squared) spin component of particle $1$
  is represented by the degenerate operator $S_{1n}^{2}\otimes I_{2}$
  on $1+2$'s Hilbert space, ontological contextualism blocks the
  conclusion (which would otherwise be forced by the Heywood-Redhead
  argument) that
  the outcome of a measurement of $S_{1n}^{2}$ must causally depend upon
 measurements performed on particle 2.  This conclusion is blocked
 because the very definition of the
 spin magnitude being measured rests on which complete commuting set of
self-adjoint
 operators of the
 composite $1+2$ system the values of $S_{1n}^{2}\otimes I_{2}$ are
referred to.  If we cannot
 specify a subsystem's properties independently of properties
 relating to the whole combined system, then the question of whether
 properties intrinsic to a subsystem causally depend on
 measurements undertaken on
 spacelike-separated systems cannot even be raised.  Michael calls
 this consequence of ontological contextualism `ontological
 nonlocality' to contrast it with `environmental
 nonlocality' that would involve an explicit spacelike causal dependence of
local
 properties on distant measurements in apparent conflict
 with relativity theory.

 Despite the lure of this route to peaceful coexistence between
 relativity and quantum nonlocality, it is hard to be totally
 at ease with an ontology that
 entertains the existence of large numbers of distinct physical
 magnitudes which are in principle
 statistically indistinguishable.  And since it's not obvious
 how failing to classify quantum nonlocality as a causal
 connection improves the chances of securing a Lorentz invariant realist
interpretation of the
 theory, it is surely worth seeking an alternative,
 \emph{simpler} realism of possessed values
 that takes the functional
 relations between self-adjoint operators just as seriously.

 In fact, one doesn't have to look very far.  The key lies in
 rejecting an assumption that is necessary to prove the Kochen-Specker theorem
 which Michael dubs the `Reality Principle' in
 \emph{Incompleteness, Nonlocality and Realism} (1987):
 \begin{quotation}
 \footnotesize{If there is an operationally defined number associated
 with the self-adjoint operator $\hat{Q}$ (i.e. distributed
 probabilistically according to the statistical algorithm of QM for
 $\hat{Q}$), then there exists an element of reality \ldots
 associated with that number and measured by it (1987, p. 133-4).}
   \end{quotation}
  Michael considers (and rightly rejects) only one way to deny the
 Reality
 Principle.  Faced with incompatible ways to measure a
 degenerate self-adjoint operator $\hat{Q}$---depending on which complete
 commuting set it is measured along with---one could say that only one
 way reveals $\hat{Q}$'s true value and the others `produce numbers
 which just ``hang in the air'' and do not measure anything of
 ontological significance' (1987, p. 136).  But this is not the most
 natural way to deny the Reality Principle.  The most natural way is
 to regard the measurement of certain self-adjoint operators as yielding
results
 without ontological significance \emph{however} they are
 `measured'---the
 paradigm example being the measurement of `spin' in Bohm's
 theory.\footnote{See Pagonis and Clifton (1995)
 and D\"{u}rr et al. (1996) for further discussion.}
 This is not at all to renounce the realist demand for an explanation of
 measurement results, but only to abandon the particular
 form of
 explanation demanded by the Reality Principle, which dictates that each
 self-adjoint operator needs to be thought of as having its
 measurement results determined by a pre-existing element of reality
 \emph{unique to that operator}.  (In Bohm's theory, by contrast, all
measurement
 results are grounded in the pre-existing position of the particle
 together with
 its initial wavefunction.) Jettisoning this part of the Reality Principle
 clears the way for
 an interpretive programme in quantum mechanics which has received concrete
expression
 recently in various `modal' interpretations of quantum
 mechanics,\footnote{See
 Clifton (1996)
 and references therein.} and has figured
 prominently in the writings of John Bell
 (1987, Chs. 5, 7, 19).

 For Bell, a self-adjoint operator is just a mathematical device
 which, when traced with the system's density operator,
 generates the empirically correct statistics in an experiment on the
 system
 which orthodox quantum mechanics would loosely call a `measurement' of the
`observable'
 represented by the operator (1987, p. 52).  Out of the `observables' of the
 orthodox interpretation Bell seeks
 to isolate some subset, the
 `\emph{be}ables' of
 a system, which
 can be ascribed determinate values and about which orthodoxy's
 loose talk is perfectly precise:
\begin{quotation}
\footnotesize{Many people must have thought along the following lines.
Could one not just promote \emph{some} of the 'observables' of the present quantum theory
to the status of beables? The beables would then be represented by linear
operators in
the state space. The values which they are allowed to \emph{be} would be
the eigenvalues
of those operators. For the general state the probability of a beable
\emph{being}
a particular value would be calculated just as was formerly calculated
the probability of \emph{observing} that value (1987, p. 41).}
\end{quotation}
Bell's thinking is the exact opposite of Michael's in \emph{From
Physics to Metaphysics}.
While Michael entertains the possibility that there are far more beables than
self-adjoint operators, Bell
is content with there being far less.  Elsewhere, Bell explains how
one can get away with this:
\begin{quotation}
\footnotesize{Not all `observables' can be given beable status, for they do
not all
have simultaneous eigenvalues, i.e. do not all commute.  It is
important to realize therefore that most of these `observables' are
entirely redundant.  What is essential is to be able to define the
positions of things, including the positions of instrument pointers
or (the modern equivalent) of ink on computer output (1987,
p. 175).}
\end{quotation}
\begin{quotation}
\noindent \footnotesize{`Observables' must be \emph{made}, somehow, out of
beables.  The
theory of local beables should contain, and give precise physical
meaning to, the algebra of local observables (1987, p. 52).}
\end{quotation}
In this last passage we see a further contrast with
Michael's thinking.  While his
ontological nonlocality countenances locally measurable but nonlocally
defined beables, Bell restricts his considerations to beables that are both
locally measurable \emph{and}
locally defined.\footnote{See also pgs. 42 and 53 of Bell's (1987).}  And
in at least
one other place (1987, p. 42), Bell again expresses his
interest in modelling a
theory of local beables
after Haag's (1992) algebras of local observables in
relativistic quantum field theory.

I find this aspect of Bell's thinking particularly
intriguing.
But given the $C^{\star}$-algebraic formulation of Haag's theory, it would
be of interest to have a purely algebraic characterization of sets of
(bounded) observables which are viable candidates for representing the
\emph{beables} of a
quantum-mechanical system---be it a spin-$1$ particle in the singlet
state or a bounded open region of
spacetime.  This is the main aim of the present paper.

I begin in Section $2.$ by considering what subsets of the self-adjoint
part of the $C^{\star}$-algebra
$\alg{U}$ of
a quantum system should be candidates for beable
status.
No doubt this is partly a matter of taste.
But a natural requirement to impose is that sets of beables be closed under
the taking of any continuous self-adjoint function of their
members.  As we shall see, such sets have their own characteristic
algebraic structure,
and I call them `Segalgebras' because they conform to the
general postulates for algebraic systems of observables laid down and
studied by
Segal (1947).
It turns out that a subset of self-adjoint elements in $\alg{U}$ forms a
Segalgebra
exactly when it is the self-adjoint part of
some $C^{\star}$-\emph{sub}algebra of $\alg{U}$.  This is satisfying insofar
as there is no reason to
expect that a set of beables should have an algebraic structure any
different from the full set of observables of the system out of which
it is distinguished.  And the fact that Segalgebras are none other than the
self-adjoint parts of $C^{\star}$-algebras allows us to carry over
facts about $C^{\star}$-algebras to their
Segalgebras.

In Section $3.$ I discuss what is going
to count as an acceptable way of assigning values to beables in a
Segalgebra.  Again this is partly a matter of taste, and has
been hotly debated ever since von Neumann proved his infamous
no-hidden-variables theorem.  Nevertheless I shall argue that
while von Neumann's
conception of values as given by linear functionals on the Segalgebra
of `observables' of a system is utterly
inappropriate if their assigned `values' are only defined dispositionally
or counterfactually
(so I agree wholeheartedly with Bell 1987, Ch. 1),
there is no reason not to require the categorically possessed
values of beables to be given by linear functionals, regardless of
whether they commute.
This will not lead in the direction of an algebraic
analogue of von Neumann's no-go theorem, such as is proved by
Misra (1967), simply because I shall not be
supposing that all `observables' have
beable status!

In line with the approach to value definiteness taken by modal
interpretations, I will also not be requiring that the beables of a quantum
system
be the same from one quantum state of the system to another.  In the second
passage from Bell
(1987) quoted above he implies that the self-adjoint operators
corresponding to beables must all commute.  It will turn out
that this only follows (at least in nonrelativistic quantum mechanics)
 if one requires the beables of a system to be the same for
all its quantum states.
However, even if that requirement is dropped, we shall see that a Segalgebra of
beables still has to be `almost
commutative'.

In Section $4.$ I shall
introduce the new notion of a quasicommutative Segalgebra to make
this idea precise. In
Section $5.$ I go on to show that quasicommutative
Segalgebras are both necessary and
sufficient for representing the measurement statistics prescribed by a
quantum state as an average over the actual values of the beables in the
algebra\footnote{This result generalizes results obtained by
Bub and Clifton (1996) and Zimba and Clifton (forthcoming)
in two
directions.  First and most importantly,
I shall not need to restrict myself to sets of beables with discrete
spectra.  And, second, I
shall not need to assume
anything about the projection operators with beable status (such as: that
they form an
ortholattice); indeed, a $C^{\star}$-algebra need have
\emph{no} nontrivial projections.}---in line with the first passage from
Bell (1987) quoted
above.
I also discuss two concrete examples of
noncommuting Segalgebras of beables employed by the orthodox (Dirac-von
Neumann) interpretation of
nonrelativistic quantum mechanics and
 modal interpretations thereof.   Section 6. then discusses the
 question of how `big' (and noncommutative) a Segalgebra of beables can be
consistent with
 satisfying the statistics of some quantum state.  There is a simple
 characterization of the Segalgebras of beables on a finite-dimensional
 Hilbert space that are maximal in this sense, but the
 infinite-dimensional case remains open.

Finally, in Section $7.$ I discuss two ways one can argue for
entertaining only commutative Segalgebras of beables.  The first way
(as I've already mentioned) is to require that the beables of a
system be the same for all its quantum states, or at least for a
`full set' thereof (we'll see that the former demand is too strong on
physical grounds).
The second way arises in the context of algebraic relativistic quantum
field theory, where it turns out that Segalgebras of local
 beables must be \emph{fully} commutative if they are to satisfy the
 measurement statistics dictated by a state of the field with bounded
 energy.

\section{Segalgebras of Beables}

A $C^{\star}$-algebra is a normed algebra
$\alg{U}$ over the complex numbers which is complete in the metric topology
induced
by the norm $|\cdot|$ and equipped with an involution $^{\star}$
that, together with the norm, satisfies the $C^{\star}$-norm
property,\footnote{Note that this property, together with the triangle
inequality and product inequality
$|AB|\leq|A||B|$, entails that $|A|=|A^{\star}|$ for any
$A\in\alg{U}$.}
\begin{equation}
\label{eq:norm}
 |A^{\star}A|=|A|^{2},\ \mbox{for all}\ A\in\alg{U}.
 \end{equation}
We will hardly ever need to suppose that our $C^{\star}$-algebras
are concrete algebras of bounded operators acting on some Hilbert
space, but for convenience I'll still refer to the elements of $\alg{U}$ as
operators.  Of course, for a quantum system represented by some $\alg{U}$,
the operators in
$\alg{U}$'s self-adjoint part---consisting of all
$A\in\alg{U}$ such that $A=A^{\star}$---represent the
bounded observables of the system.  For simplicity, I set aside the
possibility of superselection rules and take the term `observable'
to be synonomous with `self-adjoint operator'.
Our task, then, is to lay down some natural guidelines for granting
observables in $\alg{U}$
beable status.

It seems reasonable to require that a quantum system be such that
its beables combine algebraicly to yield other beables, the
idea being that if a set of observables have definite values, any
self-adjoint function of them ought to have a definite value as well.
Thus a set of beables should at least form a real vector space.
And, starting with any
single beable, one should be able to form polynomials over the
reals in that
beable which are also beables of the system.  If we are going to
allow these polynomials to have a constant term, we had better also
require that the identity operator $I$ be a beable.\footnote{Not all
$C^{\star}$-algebras
have an identity, but one can always be `adjoined' to any
$C^{\star}$-algebra---see Bratteli and Robinson 1987, Prop. 2.1.5.}
Finally, it seems
reasonable to require that sets of beables not just be closed under
polynomial functions of their members, but all continuous functions
thereof.  Thus if an observable $A$ is a beable, then $\sin{A}$,
$e^{A}$ and (if the
spectrum of A consists only of nonnegative values) $\sqrt{A}$ should
all have beable status too.  There is only one way to
define a nonpolynomial continuous function of a bounded observable $A$, viz. as
the norm limit of a sequence of polynomials in $A$ by analogy with the
Weierstrass approximation theorem from ordinary
analysis.\footnote{Geroch (1985, Ch. 52) contains a complete discussion.}
Thus sets of beables will need to
be closed in norm.\footnote{This is
consistent with our beables remaining self-adjoint, since the limit of a
sequence of self-adjoint
operators must itself be self-adjoint due to the continuity of the
adjoint operation (which follows from
$|A^{\star}-B^{\star}|=|(A-B)^{\star}|=|A-B|$).}

Of course we cannot require sets of beables to be closed under products,
since the product of two observables is an observable only if they commute.
However, we can always introduce a new
symmetric product on $\alg{U}$ by
\begin{equation}
A\circ
B\equiv 1/4[(A+B)^{2}-(A-B)^{2}]=1/2[A,B]_{+},\label{eq:symmetric}
\end{equation}
which is manifestly such that if both $A$ and $B$ are self-adjoint
$A\circ B$ will be too.  Since the symmetric product of two
self-adjoint operators is expressible, as above,
in terms of real linear combinations and squares, it follows from our
requirements on sets of beables that they \emph{are} closed under the symmetric
product.

It is easy to see that the
symmetric product on $\alg{U}$ is homogeneous (i.e. $r(A\circ
B)=(rA)\circ B=A\circ (rB)$) and distributive over addition.
Moreover, the symmetric product will be associative on any triple of
elements $A,B,C\in\alg{U}$ if they mutually commute in the $C^{\star}$ product
 (for in that case $\circ$ just reduces to the
$C^{\star}$ product, which of course is associative by definition).  However,
if a triple of elements do not mutually commute, the symmetric product
cannot be assumed associative.  A simple example is
provided by the $C^{\star}$-algebra $\alg{U}(H_{2})$ of all Hermitian
operators on complex two-dimensional Hilbert space.
If we consider the Pauli spin
operators $\sigma_{x}$ and $\sigma_{y}$, then since they anti-commute
we have
$\sigma_{x}\circ \sigma_{y}=0$, thus
$\sigma_{x}\circ(\sigma_{x}\circ\sigma_{y})=0$; yet
$(\sigma_{x}\circ\sigma_{x})\circ\sigma_{y}=\sigma_{x}^{2}\circ\sigma_{y}
=I\circ\sigma_{y}=\sigma_{y}$.

Since raising elements in $\alg{U}$ to any desired $C^{\star}$ power can be
re-expressed in
terms of the symmetric product as
\begin{equation}
 A^{n}=\underbrace{A\circ A\circ \cdots \circ
A}_{n\ \mathrm{times}}, \label{eq:power}
\end{equation}
we can dispense with reference to the $C^{\star}$ product in our
requirements on
beable sets.  Thus what we have required, so far, is that
 any set of beables be a real closed linear subspace of
 observables taken from $\alg{U}$
which forms a (not necessarily associative!) algebra with respect
to the symmetric product.  This is an instance of the sort of algebraic
structure
studied by Segal (1947), and a simple concrete example is given by
\begin{equation}
\label{eq:set}
\{a\sigma_{x}+b\sigma_{y}+cI\in \alg{U}(H_{2})|a,b,c\in\alg{R}\}.
\end{equation}

There is one last requirement I need to impose on beable sets.  We can
also introduce
on $\alg{U}$ an \emph{antisymmetric} product by
\begin{equation}
A\bullet B\equiv i/2[A,B]_{-},\label{eq:antisymmetric}
\end{equation}
which also has the property that if both $A$ and $B$ are self-adjoint
so is $A\bullet B$.  This product is again homogeneous and distributive by
not necessarily
associative (e.g.\
$\sigma_{x}\bullet(\sigma_{x}\bullet\sigma_{y})=-\sigma_{y}$, while
$(\sigma_{x}\bullet\sigma_{x})\bullet\sigma_{y}=0$).
If we are serious about wanting sets of beables to
contain \emph{all} continuous self-adjoint functions of their members,
then they ought to be closed under the antisymmetric product
too (its continuity is proven using the triangle and product
inequalities).  With closure under both the symmetric and
antisymmetric products, we then get closure under self-adjoint polynomials
in two
beables, like
\begin{equation}
cAB+c^{\ast}BA=2\Re(c)A\circ B+2\Im(c)A\bullet B.
\end{equation}
It must be admitted, though, that closure under $\bullet$ is a strong
assumption; for example, the set in
Eqn. \ref{eq:set} is now ruled out, since it fails to contain
$\sigma_{x}\bullet\sigma_{y}=-\sigma_{z}$.  It might be of
interest
to investigate what portion of my conclusions can be recovered without
assuming sets of beables are closed under $\bullet$, but I shall not
do so here.

To summarize, our candidate beable sets are to be real closed linear
subspaces of observables in $\alg{U}$ that contain the identity and are closed
under the (generally nonassociative) symmetric and antisymmetric products.
Such structures I call \emph{Segalgebras} to distinguish them within
the class
of Segal's own algebras, which need not admit an antisymmetric
product.\footnote{Neither does Segal's (1947) symmetric product (which he
calls `formal product') have to be homogeneous or
distributive!}
Virtually everything about Segalgebras follows from the fact
that they are simply the self-adjoint parts of $C^{\star}$-subalgebras of the
$C^{\star}$-algebras from which their elements are
drawn.

To see this, recall that a subalgebra of a $C^{\star}$-algebra
$\alg{U}$
is a subset of the algebra (possibly not containing the identity) that
is closed under the relevant operations, i.e. a complex norm closed
subspace of $\alg{U}$ closed under the
taking of $C^{\star}$ products and
adjoints.  For $T$ any set of observables in $\alg{U}$,
define
\begin{equation}
\label{define}
 T+iT=\{A\in\alg{U}|A=X+iY,\ \mbox{with}\ X,Y\in T\}.
 \end{equation}
 Then we have:

\begin{theorem}
\label{self-adjoint}
A subset $T$ of the observables in a $C^{\star}$-algebra $\alg{U}$ is
a real closed
linear subspace of $\alg{U}$ closed under the symmetric and
antisymmetric products
if and only if $T+iT$ is a
$C^{\star}$-subalgebra of $\alg{U}$.
\end{theorem}
\emph{Proof.}  `If'.  Assuming $T+iT$ is a subalgebra of
$\alg{U}$, it is automatic that $T$ is a real linear subspace.
Moreover, since any Cauchy sequence $\{A_{n}\}\subseteq T$ must at least
converge to an
element $A\in T+iT$, and the limit of a sequence
of self-adjoint elements must itself be self-adjoint, $A$ must lie in
the self-adjoint part of $T+iT$, which is obviously $T$.  Thus $T$ is closed.
Now recall that if $D$ is an element in a
$C^{\star}$-algebra, it has unique real and imaginery parts given by
\begin{equation}
\label{eq:parts}
\Re(D)=\frac{1}{2}(D+D^{\star}),\  \Im(D)=\frac{1}{2}(-iD+iD^{\star}).
\end{equation}
To prove $T$ is closed under symmetric and antisymmetric products,
suppose $A,B\in T$.  Then $A,B\in T+iT$, and since $T+iT$ is a
subalgebra of $\alg{U}$, $AB\in T+iT$ has unique real and imaginery parts.
Using Eqns. \ref{eq:parts}, those parts are just
$A\circ B$ and $-A\bullet B$ and the conclusion follows.

`Only if'.  Given that $T$ is a real subspace of $\alg{U}$, it is
routine to check that $T+iT$ is a complex subspace closed under $^{\star}$ .
Next, suppose $\{A_{n}\}\subseteq
T+iT$ is a Cauchy sequence, i.e.
$|A_{n}-A_{m}|\rightarrow 0$.  From Eqns. \ref{eq:parts}, the
triangle inequality and the fact that $|D|=|D^{\star}|$ for any $D\in\alg{U}$,
we see that $|\Re(D)|,|\Im(D)|\leq
|D|$.  Therefore,
\begin{equation}
\label{eq:limit}
|\Re(A_{n})-\Re(A_{m})|=|\Re(A_{n}-A_{m})|\leq |A_{n}-A_{m}|\rightarrow 0,
\end{equation}
and, similarly, $|\Im(A_{n})-\Im(A_{m})|\rightarrow 0$.
So both $\{\Re(A_{n})\}$ and $\{\Im(A_{n})\}$ must be
Cauchy sequences in $T$.  Letting their respective limits be
$A_{1},A_{2}\in \alg{T}$, further use of the triangle inequality
establishes that $A_{1}+iA_{2}$ is the
limit in $\alg{U}$ of
$\{A_{n}\}$.  Hence $T+iT$ is norm closed.  Finally, for closure of $T+iT$
under $C^{\star}$-products, let $A,B\in T+iT$, so
\begin{equation}
\label{eq:helper}
 A=X+iY\ \mbox{and}\ B=X'+iY'\ \mbox{with}\ X,Y,X',Y'\in T.
 \end{equation}
 A simple calculation yields
\begin{equation}
\label{eq:calculation1}
\Re(AB)=X\circ X'+X\bullet Y'+Y\bullet X'-Y\circ Y',
\end{equation}
\begin{equation}
\label{eq:calculation2}
\Im(AB)=-X\bullet X'+X\circ Y'+Y\circ X'+Y\bullet Y'.
\end{equation}
Therefore, since $T$ is a real linear subspace closed under both the
symmetric and
antisymmetric products, $AB\in T+iT$.\ \emph{QED}.\vspace{.15in}

Thm. \ref{self-adjoint} tells us that a subset of $\alg{U}$ is a Segalgebra
exactly when it
is the self-adjoint part of some subalgebra of $\alg{U}$ containing
the identity.  As a first example of how this makes the `theory' of Segalgebras
parasitic upon facts about the $C^{\star}$-algebras they generate,
consider the maps that preserve these structures.
Recall that a mapping of $C^{\star}$-algebras $\psi:\alg{U}\rightarrow
\alg{U}'$ is called
a $^{\star}$\emph{-homomorphism} if it preserves the identity, linear
combinations, products and
adjoints.  It is a theorem that
$^{\star}$-homomorphisms are
continuous,\footnote{Bratteli and Robinson (1987), Prop. 2.3.1.} so in fact
they preserve
\emph{all} the relevant structure of a $C^{\star}$-algebra.  Analogously,
call a mapping of Segalgebras $\phi:\alg{S}\rightarrow \alg{S}'$ a
\emph{homomorphism} if it preserves the identity, linear combinations,
and symmetric and
antisymmetric products.
There is an obvious bijective correspondence
between $^{\star}$-homomorphisms and homomorphisms.
If $\psi:\alg{U}\rightarrow \alg{U}'$
 is a $^{\star}$-homomorphism, the restriction of $\psi$ to
 $\alg{U}$'s Segalgebra is a homomorphism into $\alg{U}'$'s.
 Conversely, if $\phi:\alg{S}\rightarrow \alg{S}'$ is a homomorphism,
 the (unique) linear extension of $\phi$ to $\alg{S}+i\alg{S}$ given by
 $\psi(A)=\phi(\Re(A))+i\phi(\Im(A))$  is a $^{\star}$-homomorphism
 into $\alg{S}'+i\alg{S}'$.  (To check that $\psi$ preserves
 $C^{\star}$ products, use Eqns. \ref{eq:helper}--\ref{eq:calculation2}.)
 Due to this bijective correspondence, we learn `for free' that
homomorphisms of
 Segalgebras must also be continuous.\footnote{It is natural to ask whether
one could define Segalgebras \emph{independently}
 of $C^{\star}$-algebras (not assuming, as I have, that their elements
 are drawn from a $C^{\star}$-algebra), and then prove that every
 Segalgebra (abstractly defined)
 is isomorphic to the self-adjoint part of a
 $C^{\star}$-algebra.  (Never mind the fact that were I to have taken
 such a route, my observations about Segalgebras---such as that their
 homomorphisms are continuous---would no longer come for free!)
 I recently
 learned from Klaas
 Landsman that the answer is in fact Yes, and that the abstract counterpart
of a
 Segalgebra is called a Jordan-Lie-Banach
 algebra ($\circ$ is the Jordan product, $\bullet$ the Lie
 product)---see Landsman (forthcoming), Ch. 1 for a complete discussion and
 references.}

\section{Statistical States and Value States}

Having decided that our sets of beables will have the algebraic
structure of Segalgebras, the next step is to decide how to assign
values to beables.  For this, we first need to recall the algebraic
definition of a quantum state.

An operator $A$ in a $C^{\star}$-algebra
$\alg{U}$ is called \emph{positive} if
it is self-adjoint and has a non-negative spectrum.  It is useful to
have on hand two alternative equivalent definitions:
$A$ is positive if it is the square of a self-adjoint operator in
$\alg{U}$, or if there is a
$B\in\alg{U}$ such that
$A=B^{\star}B$.\footnote{See Bratteli and Robinson (1987), Props. 2.2.10
and 2.2.12
respectively.}
A \emph{state} on a $C^{\star}$-algebra $\alg{U}$ is a
(complex-valued) linear functional on $\alg{U}$ that maps positive operators
to nonnegative numbers and the identity to 1.  It is a theorem that states, so
defined, are
continuous.\footnote{Bratteli and Robinson (1987), Prop. 2.3.11.}

We can define a \emph{state on a
Segalgebra} $\alg{S}$ in essentially the same way, as a (this time,
real-valued) linear
functional on $\alg{S}$ that maps squares to nonnegative numbers and $I$ to
$1$.  Again, there is the obvious bijective correspondence
between states on Segalgebras and on the $C^{\star}$-algebras they
generate, so that Segalgebra states are continuous too.  In general, a
state on a Segalgebra is merely statistical and specifies
 only the expectation
values of the observables the state acts upon.  If we want to interpret
those expectation
values as averages of the actual values of beables, we need a specification
of the
allowed valuations.

Certainly a valuation on the beables in a Segalgebra
should itself be a real-valued function on it.  Moreover,
valuations ought to assign nonnegative numbers to observables with nonnegative
spectra!  And valuations should map $I$ to $1$, if only because if
some `deviant' set of valuations
did not, then since all the states whose statistics we want to
recover with our valuations map $I$ to $1$, we would have to assign
that deviant set measure-zero anyway.
But should a valuation of the beables in a
Segalgebra $\alg{S}$
be a \emph{linear} function on $\alg{S}$?  That is, should a valuation just
be a special kind of state, in the technical sense above, that gives the
actual values of beables
rather than just their expectations?  I believe that the assumption of
linearity \emph{is} defensible, yet I
completely agree with Bell's (1987, Ch. 1) critique of von Neumann's
no-hidden-variables
theorem (!).  Let me explain.

What Bell objects to in both von Neumann's
and Kochen-Specker's no-go theorems is arbitrary assumptions
about how the results of measurements undertaken with incompatible
experimental arrangements would turn out.  For von Neumann, it is the
assumption that if an observable $C$ is actually measured, where $C=A+B$
and $[A,B]\neq
0$, then had $A$ instead been measured, or $B$,
their results would have been such as to sum to the value actually
obtained for $C$.  For Kochen-Specker, who adopt von Neumann's linearity
requirement only when $[A,B]=
0$, it is the assumption
that the results of measuring $C$ would be the same independent of
whether $C$ is measured along with $A$ and $B$ or in the context of
measuring some other pair of
compatible observables $A'$ and $B'$ such that $C=A'+B'$.
What makes these assumptions arbitrary is that the results of measuring
observables $A,B,C,\ldots$ might not
reveal separate pre-existing values for them (contra Michael's Reality
Principle), but rather realize dispositions of the
system to produce those results in the context of the specific
experimental arrangements they are obtained in.  In other words, the
`observables' at issue need not all be \emph{be}ables in the
hidden-variables interpretations the no-go theorems seek to rule out.

Some commentators place undue emphasis on the fact that, in his
criticism of von Neumann's linearity assumption, Bell points out
that it is mathematically possible for the measurement results dictated by
hidden-variables in individual cases to violate linearity for
noncommuting observables even while
linearity of their expectation values is preserved
after averaging over those variables.  But it is wrong to portray Bell's
critique as turning on a mathematical possibility, and it misses the
reason why, for Bell, the theorems of von Neumann and Kochen-Specker
stand or fall together. For having established the mathematical point
beyond doubt using
a simple toy hidden-variables model, Bell goes on to remark: `At
first sight the required additivity of expectation values seems very
reasonable, and it is rather the non-additivity of allowed values
(eigenvalues) which requires explanation' (1991, p. 4).  Bell then backs up
this
remark by
giving a positive \emph{physical} explanation for the non-additivity, in terms
of measurement results displaying dispositions of the system in
different experimental contexts rather
than pre-existing values for the `observables' measured.

Now while arbitrary assumptions about the results of
measuring `observables' are certainly to be avoided, it seems to me that
there is no good physical reason (short of reintroducing some form of
ontological contextuality) to reject linearity as a
requirement on the categorically possessed values of \emph{beables}.
Of course,
linearity for the values of noncommuting beables with
discrete
spectra is not going to be easy to
satisfy, since the sum of any two eigenvalues for $A$ and $B$ needn't even
\emph{be}
an eigenvalue for $C=A+B$.\footnote{The standard example is $A=\sigma_{x},
B=\sigma_{y}$.} But rather than assign an eigenvalue to beable $C$
in such cases that
floats freely of the values assigned to beables $A$ and $B$ (yet on averaging
linearity of expectation values is miraculously restored), it would be
better not to have promoted `observables' $A$ and $B$ to beable status
in the first place!\footnote{One could avoid this conclusion by only
withholding beable status
from $C$---thereby rejecting my assumption that beable sets
 form real vector spaces.  But it would be hard to find principled
reasons for that rejection.  For example, the kinetic and potential
energies of a particle in a potential well don't commute, but if both
were assumed to have pre-existing possessed values it would be
difficult to comprehend the particle's
total energy not having a possessed value too.}
  \emph{That} is what I take the lesson of Bell's
critique of the no-hidden-variables theorems to be.

There is one last requirement to impose on our beable valuations.  The
value assigned to the square of
any beable should equal the square of
the value assigned to that beable.  It appears that Bohm's (1952) theory
contradicts this by predicting that a particle confined to a
box in a stationary state with energy $(1/2m)(nh/L)^{2}$ will possess zero
momentum, so that its energy
could not possibly be proportional to the square of the value of its
momentum.\footnote{Bohm (1952, Sec. 5).}  However, Bohm's
theory also predicts that if the momentum of the particle is measured, it will
be found to have nonzero values $\pm nh/L$ with equal probability.
It follows that momentum in the Bohm theory
is \emph{not} a beable in Bell's sense, since the probability of
\emph{finding} a certain value for momentum is not the same as the
probability that the particle \emph{has} that momentum.

So we are requiring our
beable value states to be, in the well-known jargon, dispersion-free
states.  A state $\omega$ on a Segalgebra
$\alg{S}$ is called
\emph{dispersion-free on an observable} $A\in\alg{S}$ if
$\omega(A^{2})=(\omega(A))^{2}$, and a state is called \emph{dispersion-free on}
$\alg{S}$ if
it is dispersion-free on \emph{all} observables of $\alg{S}$.  It is time
now to develop some of the consequences of our assumption that
valuations on Segalgebras of beables are given by dispersion-free
states.

Obviously
every homomorphism of $\alg{S}$ into the Segalgebra of real numbers
$\alg{R}$ is a dispersion-free state on $\alg{S}$.  (In $\alg{R}$ the
symmetric product is the usual product, the antisymmetric product of
any two real numbers is $0$, and the norm of an element is its
absolute value.)  Conversely, every dispersion-free state on $\alg{S}$ is a
homomorphism into $\alg{R}$.  In fact, somewhat more is true, as a consequence
of the following theorem.\footnote{This result is
a modification of standard arguments for
$C^{\star}$-algebras given in Kadison (1975, pp. 105-6).}

\begin{theorem}
\label{somewhatmore}
Let $\omega$ be a state on a Segalgebra $\alg{S}$ that is
dispersion-free on some $A\in\alg{S}$.  Then $\omega(A)$ lies in the
spectrum of $A$ and
for any
$B\in\alg{S}$, $\omega(A\circ B)=\omega(A)\omega(B)$ and
$\omega(A\bullet B)=0$.
\end{theorem}
\emph{Proof}.  Extend $\omega$ to a state $\tau$ on
the $C^{\star}$-algebra $\alg{S}+i\alg{S}$ by defining
$\tau(C)=\omega(\Re(C))+i\omega(\Im(C))$.  The map which sends
each pair of elements $C,D\in\alg{S}+i\alg{S}$ to $\tau(C^{\star}D)$
defines a positive semi-definite
inner product $\langle C|D\rangle$ on $\alg{S}+i\alg{S}$.  Therefore, we can
derive (in the usual way)
the Schwartz inequality
\begin{equation}
\label{eq:Schwartz}
|\tau(C^{\star}D)|\leq\sqrt{\tau(C^{\star}C)}\sqrt{\tau(D^{\star}D)},\
\mbox{for any}\ C,D\in\alg{S}+i\alg{S}.
\end{equation}

Now since $\tau$ is
dispersion-free on $A$,
$\tau([A-\tau(A)I]^{2})=0$.  Replacing $C$ by $A-\tau(A)I$ in the Schwartz
inequality above (and remembering that $A$ is self-adjoint), it follows that
$\tau([A-\tau(A)I]D)=0$ for every $D\in\alg{S}+i\alg{S}$.  Thus
$\tau(A)=\omega(A)$ must lie in the spectrum of $A$; for were
$A-\tau(A)I$ invertible in $\alg{S}+i\alg{S}$, we could set $D$ equal
to the inverse of $A-\tau(A)I$ and derive the contradiction
$\tau(I)=0$.\footnote{Since
invertibility in $\alg{S}+i\alg{S}$ is equivalent to invertibility in
any $C^{\star}$-algebra with $\alg{S}+i\alg{S}$ as a subalgebra
(Bratteli and Robinson 1987, Prop. 2.2.7), this
argument does not assume that $\alg{S}$ is the entire self-adjoint part of
the $C^{\star}$-algebra describing some system.}

Continuing, since $\tau([A-\tau(A)I]D)=0$ for every
$D\in\alg{S}+i\alg{S}$, it follows that $\tau(AD)=\tau(A)\tau(D)$ for every
such $D$.
Using the same argument, we can replace $D$ by $A-\tau(A)I$ and $C$
by $D^{\star}$ in Eqn. \ref{eq:Schwartz} to get
$\tau(DA)=\tau(D)\tau(A)$ for every $D\in\alg{S}+i\alg{S}$.  Therefore,
since $\tau$ agrees with $\omega$ on $\alg{S}$, for any
$B\in\alg{S}$ we have
\begin{eqnarray}
\omega(A\circ B) & = & \tau(1/2[AB+BA]) \\
& = & 1/2[\tau(AB)+\tau(BA)] \\
 & = & 1/2[\tau(A)\tau(B)+\tau(B)\tau(A)] \\
 & = & 1/2[\omega(A)\omega(B)+\omega(B)\omega(A)] \\
 & = & \omega(A)\omega(B)
 \end{eqnarray}
 and
 \begin{eqnarray}
\omega(A\bullet B) & = & \tau(i/2[AB-BA]) \\
& = & i/2[\tau(AB)-\tau(BA)]  \\
 & = & i/2[\tau(A)\tau(B)-\tau(B)\tau(A)]=0.
  \end{eqnarray}
\emph{QED}.\vspace{.15in}

Thm. \ref{somewhatmore} is a mixed blessing.  On the one hand, we only
required from the outset that value
states assign nonnegative values to beables with
nonnegative spectra, so it is reassuring to now see that the value
assigned to \emph{any} beable must lie \emph{within} its spectrum.  On the
other
hand,
the theorem shows that asking for a Segalgebra of noncommuting beables with
decently
behaved valuations is a tall order.  For suppose $[A,B]_{-}\neq 0$, so
that $A\bullet B\neq 0$.  Then since the spectrum of $A\bullet B$
must include $\pm |A\bullet B|$,\footnote{Cf. Geroch (1985, Thm. 60).}
$A\bullet B$ must
have at least one nonzero spectral value; yet Thm. \ref{somewhatmore} tells
us that antisymmetric products of beables
are always mapped to 0 by dispersion-free states!

It is exactly this sort of observation that leads Misra (1967, pp.
856-7)
 to conclude that hidden-variables in
algebraic quantum mechanics are impossible.   Working in the more
general context of Segal's algebras, Misra introduces the idea of a
derivation on the algebra which is a linear mapping $\textbf{D}$  from the
algebra
onto itself that satisfies the Leibniz rule with respect to the
symmetric product:
\begin{equation}
\textbf{D}(A\circ B)=A\circ \textbf{D}(B)+\textbf{D}(A)\circ B.
\end{equation}
Misra does not assume the Segal algebras he works with form the
self-adjoint parts of $C^{\star}$-algebras---that they are
\emph{Segalgebras} in
my sense---though he \emph{does} need to assume that every observable in the
algebra is the difference of two positive observables in the
algebra.\footnote{This is true for $C^{\star}$-algebras, and therefore
Segalgebras---see Prop. 2.2.11
of Bratteli and Robinson (1987).}  With that assumption,
Misra (1967, Thm. 4) shows dispersion-free states map derivations to zero: for
any $A$ in the algebra, any derivation $\textbf{D}$, and any
dispersion-free state
$\val{\cdot}$, $\val{\textbf{D}(A)}=0$.  Since the antisymmetric product of
any observable in a Segalgebra $\alg{S}$ with some given observable defines
a derivation on $\alg{S}$, Misra's result entails
Thm. \ref{somewhatmore}'s result that dispersion-free states on Segalgebras
map antisymmetric products to zero.  But, as we shall see explicitly in
the final section of this paper, Misra's conclusion that the result
`excludes hidden variables in the general algebraic setting of quantum
mechanics' (p. 857) is based upon a failure to distinguish `observables'
from beables.  Indeed, we shall shortly see that
noncommuting Segalgebras of beables have \emph{not} been excluded.

\section{Quasicommutative Segalgebras}

Clearly a Segalgebra $\alg{S}$ is \emph{commutative} if its antisymmetric
product
is trivial, i.e. $\alg{S}\bullet\alg{S}=\{0\}$.  In this section I introduce
the idea of a quasicommutative Segalgebra which we will see in the
next section
captures the precise extent to which beables can fail to
commute. To adequately motivate and characterize
`quasicommutativity' from a formal point of view, I'll first need an
alternative characterization of
commutativity and then I'll need to
discuss quotient Segalgebras.

There is a famous representation theorem for commutative $C^{\star}$-algebras
from which an analogous result for
commutative Segalgebras can be extracted.  Recall that an archetypal
example of a
$C^{\star}$-algebra is the algebra $\alg{C}(X)$ of all complex-valued
continuous
functions on a compact Hausdorff space $X$ (e.g. the interval $[0,1]$).
In $\alg{C}(X)$ linear combinations and products of functions are
defined in the obvious (pointwise) way, the adjoint of a function
is its complex conjugate, and the norm of a function is
the maximum absolute value it takes over $X$.\footnote{$\alg{C}(X)$ is closed
since the uniform limit of a sequence of (necessarily bounded) continuous
functions
on a compact $X$ is itself a continuous function on $X$---see Simmons
(1963, pp.
80-85).}  The point about
commutative $C^{\star}$-algebras is that they all arise in this way.
\emph{Every} commutative $C^{\star}$-algebra (with identity) is
$^{\star}$-isomorphic to $\alg{C}(X)$ for some
compact Hausdorff space $X$.\footnote{Bratteli and Robinson (1987, Prop.
2.1.11A).}  Since a
$^{\star}$-isomorphism of $C^{\star}$-algebras induces an
isomorphism between their Segalgebras, it follows that every commutative
Segalgebra $\alg{S}$
 is isomorphic to the
Segalgebra of all \emph{real}-valued continuous functions $\alg{C}(X)$ on
some compact
Hausdorff space $X$.\footnote{See also Segal (1947, Thm. 1).}

Now call a set of states $\Omega$ on a Segalgebra $\alg{S}$ \emph{full}
if $\Omega$ `separates the points of $\alg{S}$' in the sense that for any two
distinct elements of $\alg{S}$ there is a state in $\Omega$ mapping them to
different expectation values---or, equivalently (by the linearity of
states), if
for any nonzero $A\in \alg{S}$ there is a state $\omega\in\Omega$
such that $\omega(A)\neq0$.  Then the alternative characterization of
commutativity I need is the following.\footnote{This result is
just a variation on Segal's (1947, Thm. 3).}

\begin{theorem}
\label{commutative}
A Segalgebra is commutative if and only if it has a full set
of dispersion-free states.
\end{theorem}
\emph{Proof}.  `Only if'.  If $\alg{S}$ is
commutative it is isomorphic to the set of real-valued functions
$\alg{C}(X)$ on some compact Hausdorff space $X$.  Suppose $A\neq0$.
Then
$A(x)$, the isomorphic image of $A$ in $\alg{C}(X)$,
cannot be the zero function (since the isomorphism can only map the
zero operator to that
function).  So there is at least one point $x_{0}\in X$
such that $A(x_{0})\neq0$.  It is easy to see that the map
defined by $\val{B}=
B(x_{0})$ for all $B\in\alg{S}$ is a dispersion-free state on $\alg{S}$
satisfying $\val{A}\neq0$.

`If'.  Consider any pair $A,B\in\alg{S}$ and their antisymmetric product
$A\bullet B$.  If $A\bullet B\neq0$, then by hypothesis there is a
dispersion-free state $\val{\cdot}$ on $\alg{S}$ such that $\val{A\bullet
B}\neq0$.  But this contradicts Thm. \ref{somewhatmore}, therefore
$A\bullet B=0$.\ \emph{QED}.\vspace{.15in}

\noindent In light of this result, the natural way to define
quasicommutivity of a Segalgebra
is in terms of it admitting a `nearly full' set of dispersion-free states
that only separate the points of $\alg{S}$
 modulo some
ideal in the algebra.  It will turn out that factoring out the ideal
yields a quotient Segalgebra
that is (fully)
commutative---which is what one might expect if the original
(unfactored) Segalgebra were already `close to being commutative'.

I'll introduce the idea of ideals and quotients for Segalgebras by
again recalling their
$C^{\star}$-algebra counterparts first.
A (two-sided) ideal in $\alg{U}$ is a subspace $\alg{I}$ of $\alg{U}$
which is
invariant under multiplication on the left or right by any $A\in
\alg{U}$, i.e.
$\alg{U}\alg{I}\subseteq\alg{I}$ and
$\alg{I}\alg{U}\subseteq\alg{I}$.  We shall be interested only in closed
$^{\star}$-ideals, i.e. ideals closed in norm and under the taking of
adjoints. (An example is the collection of all functions in $\alg{C}(X)$
which vanish at some point of $X$.)  Clearly such an ideal is a subalgebra
of $\alg{U}$.  The
idea is to `factor out' this subalgebra so that what remains is again
a $C^{\star}$-algebra.  So as not to be left with something
completely trivial, we will also require that $\alg{I}$ be \emph{proper},
i.e. that
it be a
proper subset of $\alg{S}$, which is equivalent to requiring that
$\alg{I}$ not contain the identity.

Each proper closed
$^{\star}$-ideal $\alg{I}$ in
$\alg{U}$ determines an equivalence relation $\cong_{\alg{I}}$ on
$\alg{U}$ defined by
\begin{equation}
A\cong_{\alg{I}}B\ \mbox{if and only if}\ A-B\in \alg{I}.
\end{equation}
The equivalence classes of $\cong_{\alg{I}}$ form a $C^{\star}$-algebra called
the \emph{quotient $C^{\star}$-algebra} $\alg{U}/\alg{I}$ \emph{by the
ideal} $\alg{I}$.  To see how,
let $\hat{A}$ denote the equivalence class in which $A$ lies, and
similarly for $\hat{B}$, $\hat{C}$, etc.  Now define the relevant
operations in $\alg{U}/\alg{I}$ by
\begin{equation}
\label{eq:quotient}
c\hat{A}=\widehat{cA},\  \hat{A}+\hat{B}=\widehat{A+B},\
\hat{A}\hat{B}=\widehat{AB},\  \hat{A}^{\star}=\widehat{A^{\star}}.
\end{equation}
noting that $\hat{I}$ is the identity in $\alg{U}/\alg{I}$ and $\hat{0}$
the zero element.  (The factoring has been implemented by `collapsing'
everything in $\alg{I}$ into $\hat{0}\in \alg{U}/\alg{I}$.)  Because
$\alg{I}$ is a $^{\star}$-ideal, the
definitions in Eqn. \ref{eq:quotient} are well-defined independent of the
representatives chosen for the equivalence classes that appear in
them.   (For example,
$\hat{A}\hat{B}=\widehat{AB}$ is well-defined
only if $A\cong_{\alg{I}}A'$ and $B\cong_{\alg{I}}B'$ imply
$AB\cong_{\alg{I}}A'B'$.  But assuming the former, $(A-A')(B+B')\in
\alg{I}$ and $(A+A')(B-B')\in
\alg{I}$, therefore
\begin{equation}
\label{well-defined}
 \frac{1}{2}[(A-A')(B+B')+(A+A')(B-B')]=AB-A'B'\in
\alg{I}.)
\end{equation}
If we define
\begin{equation}
|\hat{A}|=\inf_{B\cong_{I}A}|B|,
\end{equation}
an elementary argument in the theory
of Banach algebras establishes
that $\alg{U}/\alg{I}$ is a complete normed algebra, and a not so
elementary argument establishes the
$C^{\star}$-norm property.\footnote{See Simmons (1963, Thm. 69D) and
 Bratteli and Robinson (1987, Prop. 2.2.19), respectively.}

The analogue of all this for Segalgebras may now be obvious.
A proper closed ideal in $\alg{S}$ is a closed subspace $\alg{I}$ of $\alg{S}$
not containing the identity which is
invariant under symmetric and antisymmetric multiplication, i.e.
$\alg{S}\circ\alg{I}\subseteq\alg{I}$ and
$\alg{S}\bullet\alg{I}\subseteq\alg{I}$.
  (Clearly we do not need to
distinguish left from right multiplication.)  By an argument
virtually identical to the `Only if' part of Thm. \ref{self-adjoint}'s
proof, $\alg{I}$ extends to a proper closed
$^{\star}$-ideal $\alg{I}+i\alg{I}$ in $\alg{S}+i\alg{S}$.  So we can
take the \emph{quotient Segalgebra} $\alg{S}/\alg{I}$ \emph{by the
ideal} $\alg{I}$ to be the Segalgebra part of
$(\alg{S}+i\alg{S})/(\alg{I}+i\alg{I})$.
It is easy to see that for $A\in\alg{S}$, $\hat{A}\in\alg{S}/\alg{I}$
and that $A,B\in\alg{S}$
lie in the same equivalence class of $\alg{S}/\alg{I}$, i.e.
$\hat{A}=\hat{B}$, if and
only if $A\cong_{\alg{I}}B$.  Furthermore, using Eqns.
\ref{eq:quotient} it is easy to verify that the map which sends
$A\in\alg{S}$ to $\hat{A}\in\alg{S}/\alg{I}$ is
a homomorphism, and this homomorphism is surjective since
$\hat{A}\in\alg{S}/\alg{I}$ is the image of
$\Re(A)\in\alg{S}$.

We have now assembled all the necessary machinery to fulfill my
promise to introduce a reasonable notion of quasicommutativity.

\begin{theorem}
\label{quasicommutative}
Let $\alg{S}$ be a Segalgebra with proper closed ideal $\alg{I}$.  Then the
following are equivalent:
\begin{enumerate}
\item $\alg{S}/\alg{I}$ is commutative.
\item For any $A\not\in\alg{I}$ there is a dispersion-free state
$\val{\cdot}$ on $\alg{S}$ such that $\val{A}\neq0.$
\item $\alg{S}\bullet\alg{S}\subseteq\alg{I}.$
\end{enumerate}
\end{theorem}
\emph{Proof}.  $1.\Rightarrow 2.$  If $A\not\in\alg{I}$, then
$\hat{A}\neq\hat{0}$.  Since $\alg{S}/\alg{I}$ is commutative (by
hypothesis), it has a full set of
dispersion-free states (Thm. \ref{commutative}).  So there is a
dispersion-free state
$\{\cdot\}$ on $\alg{S}/\alg{I}$ such
that $\{\hat{A}\}\neq0$.  Defining $\val{B}=
\{\hat{B}\}$ for all $B\in\alg{S}$, the map $\val{\cdot}$---being the
composition two
homomorphisms, the second into $\alg{R}$---is a
dispersion-free state on $\alg{S}$ satisfying $\val{A}\neq0$.

$2.\Rightarrow 3.$  Identical to the `If' part of Thm.
\ref{commutative} with
$\alg{I}$ now playing the role of $\{0\}$.

$3.\Rightarrow 1.$  Let $\hat{A},\hat{B}\in\alg{S}/\alg{I}$.  Since
$\alg{S}\bullet\alg{S}\subseteq\alg{I}$ (by hypothesis) and
$\hat{\alg{I}}=\hat{0}$,
we have $\hat{A}\bullet\hat{B}=\widehat{A\bullet B}=\hat{0}$,
whence $\alg{S}/\alg{I}$ is commutative.\ \emph{QED}.\vspace{.15in}

\noindent In the case $\alg{I}=\{0\}$, Parts $1.$ and $3.$ of Thm.
\ref{quasicommutative}
just assert that $\alg{S}$ itself is commutative and Part $2.$ that
$\alg{S}$ has a full set of dispersion-free states.  So Thm.
\ref{quasicommutative}
generalizes Thm. \ref{commutative} by relaxing its requirement of full
commutativity.
Motivated by this, call a Segalgebra $\alg{I}$-\emph{quasicommutative}
whenever it satisfies the equivalent conditions of Thm.
\ref{quasicommutative}.\footnote{This idea
deliberately parallels the idea of a `quasidistributive
lattice' introduced in Bell and Clifton (1995, Thm. 1).  I should also note
that
results similar to $1.\Leftrightarrow 2.$ of Thm.
\ref{quasicommutative} have been
proven by Misra (1967, Thm. 1)  and Plymen (1968, Thm. 4.2)
 in the context of $C^{\star}$- and von Neumann algebras,
 respectively.}
Notice from 3. of Thm.
\ref{quasicommutative} that if $\alg{S}$
is $\alg{I}$-quasicommutative and $\alg{J}$ is another proper ideal in
$\alg{S}$ containing $\alg{I}$, then $\alg{S}$ is
$\alg{J}$-quasicommutative as well.  In particular, if $\alg{S}$ is
commutative, it is automatically
$\alg{I}$-quasicommutative with respect to any proper ideal
$\alg{I}\in\alg{S}$.  Since the converse fails, quasicommutativity is
genuinely weaker than commutativity.

\section{Beable Subalgebras}

It is high time I spelled out the connection between quasicommutativity
and the problem of
\emph{beables} in quantum
mechanics.  For a Segalgebra of beables to satisfy the
statistics prescribed by some quantum state, we must be able to
interpret the state's expectation values as averages over the actual values of
the beables in the algebra.  In other words, the quantum state must be
a mixture of dispersion-free states on the algebra.

Let $x$ be a variable in a measure space $X$, $\mu$ a positive
measure on $X$ such that $\mu(X)=1$, and $\omega_{x}$ ($x\in X$) a
collection of states on a
Segalgebra $\alg{S}$.  Then the mapping defined by
\begin{equation}
\omega(A)=\int_{X} \omega_{x}(A)\mbox{d}\mu(x),\ \mbox{for any}\
A\in\alg{S},  \label{eq:int}
\end{equation}
will also be a state on $\alg{S}$.  A state $\omega$ is called
\emph{mixed} if it can be represented, in the above way, as a weighted
average of two or more (distinct) states with respect to some
positive normalized measure; if it cannot, then $\omega$ is called
\emph{pure}.  A subalgebra
$\alg{B}$ of $\alg{S}$ will be said to \emph{have beable status for
the state}
$\omega$ if $\omega|_{\alg{B}}$---the restriction of $\omega$ to
$\alg{B}$---is either a mixture of dispersion-free states on $\alg{B}$
or itself dispersion-free. As a check on the adequacy of this
definition, we get the
following
intuitively expected result.\footnote{Here, I follow Segal's (1947, p.
933) argument
almost to the letter.}

\begin{theorem}
\label{every}
Every commutative subalgebra of $\alg{S}$ has beable status for
every state on $\alg{S}$.
\end{theorem}
\emph{Proof}.  Let $\alg{C}$ be a commutative subalgebra of $\alg{S}$
and $\omega$ any of $\alg{S}$'s states.  Since $\alg{C}$ is
commutative, it is isomorphic to the set of real-valued functions
$\alg{C}(X)$ on some compact Hausdorff space $X$.  Defining
$\phi(A(x))=\omega(A)$ for every $A\in\alg{C}$,
$\phi$ is a state on $\alg{C}(X)$.  By the Riesz-Markov
representation theorem,\footnote{See Rudin (1974, Thm. 2.14).} $\phi$ must
take the
form
\begin{equation}
\phi(A(x))=\int_{X} A(x)\mbox{d}\mu_{\phi}(x),\ \mbox{for any}\
A(x)\in\alg{C}(X),  \label{eq:riesz}
\end{equation}
where $\mu_{\phi}$ is some positive normalized (completely additive) measure
on $X$.  But since for any $x\in X$ the map
$\val{A}_{x}=A(x)$ defines a dispersion-free state
on $\alg{C}$, Eqn. \ref{eq:riesz} exhibits $\omega|_{\alg{C}}$ as a
mixture of dispersion-free states on $\alg{C}$ (if
$\omega|_{\alg{C}}$ is not already one of those states itself, which
would correspond to the case where the complement of some point in
$X$ has $\mu_{\phi}$-measure zero). \ \emph{QED}.\vspace{.15in}

At this point, it is instructive to look at Bohm's (1952) theory which
supplies a concrete example of a commutative subalgebra with beable status for
every state.  For simplicity, consider the space of states of a single spinless
particle in one-dimension given by the
Hilbert space $L_{2}(\alg{R})$ of all (measurable) square-integrable,
complex-valued
functions on $\alg{R}$. The position $\hat{x}$ of the
particle and all self-adjoint functions thereof are the only true beables
in Bohm's
theory.  Of course, Segalgebras cannot contain unbounded observables.
But any assignment of a value to some unbounded self-adjoint operator
which is a function of
position, such as $\hat{x}$ itself, is equivalent to assigning a corresponding
set of values to self-adjoint operators which \emph{are} bounded functions
of $\hat{x}$, such as its spectral
projections.  Thus, we lose no generality by characterizing Bohm's
theory as granting beable status to all \emph{bounded} self-adjoint
operator-valued functions of
$\hat{x}$.

Such functions form a commutative
Segalgebra with beable status for every wavefunction in
$L_{2}(\alg{R})$.  To see this, let $f$ be any (measurable) essentially
bounded,
complex-valued function on $\alg{R}$ and define the bounded self-adjoint
operator $\hat{O}_{f}$ by
\begin{equation}
\hat{O}_{f}(\psi(x))=f(x)\psi(x)\ \mbox{for each}\ \psi(x)\in L_{2}(\alg{R}).
\end{equation}
The set of all such operators is obviously commutative,
and it is well-known that they
form a $C^{\star}$-subalgebra of the $C^{\star}$-algebra of all
bounded operators on $L_{2}(\alg{R})$.\footnote{See Geroch (1985,
Ch. 49) and Kadison and Ringrose (1983, Sec. 4.1).}  If $f$ is some bounded
function
of $x$, $\hat{O}_{f}$ is the corresponding operator-valued function of
$\hat{x}$.
In fact, the operators $\{\hat{O}_{f}\}$ capture all the
bounded operators which are functions of $\hat{x}$, since
any such function would have to commute with all the $\hat{O}_{f}$'s,
and it is known that they form a maximal commutative set of bounded
operators on $L_{2}(\alg{R})$.\footnote{Kadison and Ringrose (1983,
p. 308).}  The subset of $\{\hat{O}_{f}\}$ where the $f$'s are
real-valued functions (almost everywhere) is therefore the commutative
sub-Segalgebra $\alg{S}_{\hat{x}}$ of
all bounded observables that are functions of $\hat{x}$.
(In particular, the spectral
projections of $\hat{x}$ correspond to characteristic functions
whose characteristic sets in $\alg{R}$ have nonzero measure.)

Now for every $\hat{O}_{f}\in\alg{S}_{\hat{x}}$, define
$\delta_{r}(\hat{O}_{f})=f(r)$---so $\delta_{r}$ is the Dirac delta
distribution
at the point
$r$.  Clearly $\delta_{r}$ qualifies as a state on
$\alg{S}_{\hat{x}}$ (note how liberal the algebraic definition of a
state is!)
and is dispersion-free.  Furthermore, for any $\psi(x)\in L_{2}(\alg{R})$
and any $\hat{O}_{f}\in\alg{S}_{\hat{x}}$
we have
\begin{equation}
\int_{\alg{R}} \psi^{\ast}(x)\hat{O}_{f}\psi(x)dx=\int_{\alg{R}}
f(x)|\psi(x)|^{2}dx=\int_{\alg{R}}
\delta_{x}(\hat{O}_{f})d\rho_{\psi}(x),
\end{equation}
so that every state of $L_{2}(\alg{R})$ is indeed a mixture of dispersion-free
(Dirac) states on $\alg{S}_{\hat{x}}$.

To show that it is not necessary that subalgebras with
beable status for some state be commutative, I need the following
result about the ideals determined by states on
Segalgebras.\footnote{Once again, this result just adapts
standard $C^{\star}$-algebraic arguments to the present Segalgebraic
context---see
Kadison (1975, pp. 103-4).}

\begin{theorem}
\label{somewhatmore2}
If $\omega$ is a state on a Segalgebra $\alg{S}$, the set
\begin{equation}
\alg{I}_{\omega}=\{A\in\alg{S}|\omega(A^{2})=0\}
\end{equation}
 is a proper closed ideal in $\alg{S}$
 on which $\omega$ is dispersion-free.
 \end{theorem}
\emph{Proof}.  As in the proof of Thm. \ref{somewhatmore}, extend $\omega$
to a state $\tau$ on
the $C^{\star}$-algebra $\alg{S}+i\alg{S}$.  Fix an arbitrary
$A\in\alg{I}_{\omega}$, so $\omega(A^{2})=0$.  Replacing $C$
by $A$ in the Schwartz inequality (Eqn. \ref{eq:Schwartz}) yields
$\tau(AD)=0$ for all $D\in\alg{S}+i\alg{S}$.  In particular, with
$D$ replaced by $I$ we see that $\tau(A)=\omega(A)=0$, so that $\omega$ is
dispersion-free on the set $\alg{I}_{\omega}$.

Now let $B$ be any element of $\alg{S}$.  We need to show that
both $A\circ B$ and $A\bullet B$ lie in $\alg{I}_{\omega}$.
With $D$ replaced by $B^{2}A$ in the argument above,
$\tau(AB^{2}A)=0$ and therefore $\omega(\Re(AB^{2}A))=0$.
Using Eqns. \ref{eq:parts} and the properties of the symmetric and
antisymmetric product,
\begin{eqnarray}
\Re([AB][BA])&=&[AB]\circ[BA] \\
&=&[A\circ B-i(A\bullet B)]\circ[B\circ A-i(B\bullet A)] \\
&=&[A\circ B-i(A\bullet B)]\circ[A\circ B+i(A\bullet B)] \\
&=&(A\circ B)^{2}+(A\bullet B)^{2}.
\end{eqnarray}
Therefore since $\omega$ acts positively on squares, $\omega((A\circ
B)^{2})=\omega((A\bullet
B)^{2})=0$, as required.

Finally, note that $\alg{I}_{\omega}$ \emph{is} a real closed
linear subspace of $\alg{S}$ not containing the identity.  For
example, to get closure under vector sums,
assuming $A,B\in\alg{I}_{\omega}$ implies
\begin{eqnarray}
\label{eq:closed1}
\omega([A+B]^{2}) & = & \omega([A+B]\circ [A+B])  \\
\label{eq:closed2} & =& \omega(A^{2})+\omega(B^{2})+2\omega(A\circ
B) \\
\label{eq:closed3} & = & 2\omega(A)\omega(B) =0
\end{eqnarray}
using the fact that $\omega$ is dispersion-free on $A$ and $B$ and
Thm. \ref{somewhatmore}.\ \emph{QED}.\vspace{.15in}

If one wants to include some pair of noncommuting observables $A$ and $B$ in a
subalgebra with beable status
for some state $\omega$, the `trick' is simply to make sure $A\bullet B$
lies inside
the ideal $\alg{I}_{\omega}$ determined by that state.  If so, then it won't
matter that dispersion-free states map antisymmetric products of
noncommuting observables to
zero, because $A\bullet B$ will \emph{also} get assigned value zero in the
state $\omega$!
In short, to have
beable status in some state it is enough for a subalgebra to be
quasicommutative
with respect to the ideal determined by the state.  It is also
necessary, as shown by the following theorem.

 \begin{theorem}
\label{defsubalg}
Let $\alg{B}$ be a subalgebra of a Segalgebra $\alg{S}$ and $\omega$ a state on
$\alg{S}$.   Then $\alg{B}$ has beable status for $\omega$ if and only if
$\alg{B}$ is $\alg{I}_{\omega|_{\alg{B}}}$-quasicommutative.
\end{theorem}
\emph{Proof}.  `Only if'.  $\alg{B}$'s
$\alg{I}_{\omega|_{\alg{B}}}$-quasicommutativity is easily inferred
from
$2.$ of Thm. \ref{quasicommutative}.  Thus if $A\in\alg{B}$ with
$A\not\in\alg{I}_{\omega|_{\alg{B}}}$, then $\omega|_{\alg{B}}(A^{2})\neq0$.
But by hypothesis, there is a
collection of dispersion-free states $\{\val{\cdot}_{x}|x\in X\}$ on
$\alg{B}$ of which $\omega|_{\alg{B}}$ is
a mixture.  Therefore, for at least one $x_{0}\in X$, $\val{A^{2}}_{x_{0}}\neq
0$ and $\val{A}_{x_{0}}\neq
0$ as required.

`If'.  By 1. of Thm. \ref{quasicommutative},
$\alg{B}/\alg{I}_{\omega|_{\alg{B}}}$ is commutative.  Define the map
\begin{equation}
\phi:\alg{B}/\alg{I}_{\omega|_{\alg{B}}}\rightarrow\alg{R}\ \mbox{by}\
\phi(\hat{A})=\omega|_{\alg{B}}(A). \label{eq:pain}
\end{equation}
Since the `hat' map is surjective, this defines $\phi$ on all of
$\alg{B}/\alg{I}_{\omega|_{\alg{B}}}$, and it is easy to check that $\phi$ is
indeed \emph{well}-defined.
Furthermore, since the hat map is a
homomorphism, $\phi$ is a state on $\alg{B}/\alg{I}_{\omega|_{\alg{B}}}$.
By the commutativity of $\alg{B}/\alg{I}_{\omega|_{\alg{B}}}$ and Thm.
\ref{every},
$\phi$ is a mixture of dispersion-free
states $\val{\cdot}_{x}$ ($x\in X$) on
$\alg{B}/\alg{I}_{\omega|_{\alg{B}}}$.  But
for any $\val{\cdot}_{x}$ on $\alg{B}/\alg{I}_{\omega|_{\alg{B}}}$,
$\val{\widehat{\cdot}}_{x}$ is a dispersion-free state on $\alg{B}$.  So since
$\omega|_{\alg{B}}(\cdot)=\phi(\widehat{\cdot})$ (Eqn. \ref{eq:pain}),
$\omega|_{\alg{B}}$ is a mixture of
dispersion-free states $\val{\widehat{\cdot}}_{x}$ ($x\in X$) on
$\alg{B}$.\ \emph{QED}.\vspace{.15in}

The examples of noncommutative subalgebras of beables that I shall consider
make use of
the following result.

\begin{theorem}
\label{defset}
For any state $\omega$ on a Segalgebra $\alg{S}$, \emph{the definite set
of} $\omega$ defined by
\begin{equation}
\label{eq:defset}
\alg{D}_{\omega}=\{A\in\alg{S}|\omega(A^{2})=(\omega(A))^{2}\}
\end{equation}
has beable status for $\omega$.
\end{theorem}
\emph{Proof}.  To see that $\alg{D}_{\omega}$ is a subalgebra
of $\alg{S}$, it is easiest to use the
fact that $A\in\alg{D}_{\omega}$ if and only if
$A-\omega(A)I\in\alg{I}_{\omega}$ and invoke the properties of closed
ideals.  To illustrate, let $A,B\in\alg{D}_{\omega}$.  Then $(A-\omega(A)I)\circ(B+\omega(B)I)\in
\alg{I}_{\omega}$ and $(A+\omega(A)I)\circ(B-\omega(B)I)\in
\alg{I}_{\omega}$, thus
\begin{eqnarray}
 \frac{1}{2}[(A-\omega(A)I)\circ(B+\omega(B)I)+(A+\omega(A)I)\circ(B-\omega(
B)I)] \\
 = A\circ B-\omega(A)\omega(B)I\in\alg{I}_{\omega}.
\end{eqnarray}
But since $\omega(A)\omega(B)=\omega(A\circ B)$ by Thm.
\ref{somewhatmore}, this means $A\circ B\in\alg{D}_{\omega}$.

The beable status of $\alg{D}_{\omega}$ for $\omega$
follows trivially from $2.$ of Thm. \ref{quasicommutative} and Thm.
\ref{defsubalg}.
Thus if $A$ lies in $\alg{D}_{\omega}$ but not in $\alg{I}_{\omega}$,
then of course there \emph{is} a dispersion-free state on $\alg{D}_{\omega}$
mapping $A$ to a nonzero value---$\omega$ itself!   \
\emph{QED}.\vspace{.15in}

\noindent In the case where $\omega$ is represented by a state vector
$|v\rangle$ in a Hilbert space representation of the Segalgebra
$\alg{S}$ (so $\omega(A)=\langle v|A|v\rangle$ for all $A\in\alg{S}$),
$\omega$'s definite set consists of all those
(bounded) self-adjoint operators on the Hilbert space that share the
eigenstate $|v\rangle$---a highly noncommutative set if the space has
more than two dimensions.  This is the orthodox (Dirac-von Neumann)
`eigenstate-eigenvalue link' approach to assigning definite values to
observables.

Definite sets can be used to build subalgebras with beable status for
a state $\omega$ that are \emph{not} just subalgebras of $\omega$'s own
definite set.  The next result specifies a general class of examples
of this sort, containing Thm. \ref{defset} as a degenerate case (when
$\omega=\omega_{x}$
for all $x\in X$).

\begin{theorem}
\label{generic}
Let $\alg{S}$ be a Segalgebra, $\omega$ a state on $\alg{S}$ and
$\omega_{x}$ ($x\in X$) any family of states satisfying $\bigcap_{x\in
X}\alg{I}_{\omega_{x}}\subseteq\alg{I}_{\omega}$.  Then
$\alg{B}_{\{\omega_{x}\}}=\bigcap_{x\in
X}\alg{D}_{\omega_{x}}$
has beable status for $\omega$.
\end{theorem}
\emph{Proof}.  Since $\alg{B}_{\{\omega_{x}\}}$ is the intersection of
a collection of subalgebras of $\alg{S}$ (Thm. \ref{defset}), it is itself
a subalgebra.  To establish beable status for $\omega$, all we need
to show (by $3.$ of Thm. \ref{quasicommutative} and Thm.
\ref{defsubalg}) is that
$\alg{B}_{\{\omega_{x}\}}\bullet\alg{B}_{\{\omega_{x}\}}\subseteq\alg{I}_{\o
mega}$.
So suppose $A,B\in\alg{B}_{\{\omega_{x}\}}$.  Then both $A$ and $B$ lie
in the definite sets of all the $\omega_{x}$'s.  This is
equivalent to both
$A-\omega_{x}(A)I$ and $B-\omega_{x}(B)I$ lying in $\alg{I}_{\omega_{x}}$
for all
$x\in X$, which
implies
\begin{equation}
(A-\omega_{x}(A)I)\bullet(B-\omega_{x}(B)I)=A\bullet B\in\alg{I}_{\omega_{x}}
\end{equation}
for all $x\in X$.  But by hypothesis, $\bigcap_{x\in
X}\alg{I}_{\omega_{x}}\subseteq\alg{I}_{\omega}$,
therefore $A\bullet B\subseteq\alg{I}_{\omega}$.\ \emph{QED}.\vspace{.15in}

\noindent To get a concrete example of Thm. \ref{generic}, consider again
the case where $\omega$ is represented by a state vector
$|v\rangle$ in a Hilbert space representation $H$.
Let $R$ be any bounded self-adjoint operator on $H$ with discrete
spectrum and eigenprojections $\{R_{i}\}$ (counting only those for
 which $\omega(R_{i})>0$), and consider the (renormalized) orthogonal
 projections $\{|v_{R_{i}}\rangle\}$ of the state vector $|v\rangle$ onto
 the eigenspaces of $R$.  Since any observable with eigenvalue $0$ in all
 the states $\{|v_{R_{i}}\rangle\}$ must have eigenvalue $0$ in the state
 $|v\rangle$ (the latter lying in the span of the former), the conditions
of Thm. \ref{generic} are
 satisfied so that $\alg{B}_{\{|v_{R_{i}}\rangle\}}$ has beable status
 in the state
 $|v\rangle$.  This is the modal (i.e. state-dependent)
 method adopted by Bub (1997)\footnote{See also Bub and Clifton (1996).}
for building a set of beables out
 of the state of a system and a `preferred observable'
 $R$.\footnote{Bub's method
 turns out to include the Kochen-Dieks modal interpretation
 (discussed in Clifton 1995) as a
 special case---see Bub 1997, Sec. 6.3 for the argument.}
 It is easy to see that a `Bub-definite' subalgebra
 $\alg{B}_{\{|v_{R_{i}}\rangle\}}$ will not be a subalgebra of
 $\alg{D}_{|v\rangle}$ unless $|v\rangle$ is an eigenstate of $R$; in
 fact,
 the Bub-definite subalgebra in that case \emph{coincides} with the
 definite set of $|v\rangle$.
 Also Bub-definite subalgebras
 will not generally be commutative: the main exceptions are when $H$
  is two-dimensional, and when $R$ is nondegenerate with
  all its eigenvalues given nonzero probability by $|v\rangle$.

 \section{Maximal Beable Subalgebras}

 Bub-definite subalgebras have the extra feature that
 their projections form a maximal determinate sublattice of the
 ortholattice of projections on $H$.  This means that any enlargement
 of the projection lattice of a Bub-definite subalgebra
$\alg{B}_{\{|v_{R_{i}}\rangle\}}$
 generates an ortholattice from which it is impossible
 to recover the probabilities prescribed by
 $|v\rangle$ as a measure over the two-valued homomorphisms on the
 enlarged lattice.\footnote{Bub (1997, Sec. 4.3).}  Since I have stopped
short of making
 any a priori assumptions
 about the lattice structure of the projections in Segalgebras, it is
 natural to ask whether Bub-definite
 subalgebras are still maximal \emph{as Segalgebras}. Indeed, the
 general question of
 when (if ever) Segalgebras of beables are `as big as one can possibly get' for
 a given state is philosophically interesting in its own right, since
 an answer would seem to set a limit on how far a simple realism of
 possessed values in quantum
 mechanics can be pushed.

 Call a subalgebra
 $\alg{B}$ of $\alg{S}$
 with beable status for a state $\omega$ a \emph{maximal beable
 subalgebra for} $\omega$ if it is not properly contained in any other
 subalgebra of $\alg{S}$ with beable status for $\omega$.  An easy
application of
 Zorn's lemma shows that a maximal beable subalgebra for any given
 state always exists.  The following result
 gives an explicit (but not completely general) characterization of
 maximal beable subalgebras that covers the case of
 Bub-definite subalgebras.

  \begin{theorem}
 \label{vectormaximal}
 Let $\alg{S}(H)$ be the Segalgebra of all bounded self-adjoint operators on a
 Hilbert space $H$ and let $|v\rangle$ be any state vector in $H$.  Then if
$|v_{x}\rangle$ ($x\in X$) is any family of
vectors in $H$ satisfying:
\begin{enumerate}
\item the members of $\{|v_{x}\rangle\}$ are mutually orthonormal,
\item $|v\rangle$ is in the closed span of $\{|v_{x}\rangle\}$, and
\item $|v\rangle$ is not orthogonal to any member of $\{|v_{x}\rangle\}$,
\end{enumerate}
$\alg{B}_{\{|v_{x}\rangle\}}$ is a
maximal beable subalgebra of $\alg{S}(H)$ for the
state $|v\rangle$.  If $H$ is finite-dimensional, the converse also
holds, i.e. if
$\alg{B}\subseteq\alg{S}(H_{n})$
is a maximal beable subalgebra for $|v\rangle\in H_{n}$, then there exists a
family of vectors $|v_{x}\rangle$ ($x\in X$) in $H_{n}$
satisfying $1.$--$3.$ such that $\alg{B}=\alg{B}_{\{|v_{x}\rangle\}}$
(and this family is unique up to phases).
 \end{theorem}
 \emph{Proof}.  For the first claim, assuming $2.$
 the beable status of $\alg{B}_{\{|v_{x}\rangle\}}$ for
 $|v\rangle$ follows exactly as discussed for Bub-definite
 subalgebras at the end of the last section.  So all that remains is to
prove maximality.
 Note first that because the members of $\{|v_{x}\rangle\}$ are mutually
 orthogonal by $1.$, each is an eigenvector (with eigenvalue $0$ or $1$)
 of all the
 one-dimensional projection
 operators $P_{|v_{x}\rangle}$ $(x\in X)$.
 Consequently, the set $\{P_{|v_{x}\rangle}\}$ is contained in
 $\alg{B}_{\{|v_{x}\rangle\}}$.  Now consider the subalgebra
 $\alg{T}$ generated by $\alg{B}_{\{|v_{x}\rangle\}}$ and any
 $A\not\in\alg{B}_{\{|v_{x}\rangle\}}$.  Our task is to show that
 $\alg{T}$ cannot have beable status for $|v\rangle$---so suppose (for reductio
 ad absurdem) that $\alg{T}$ does.  Then for any $x\in X$, $A\bullet
 P_{|v_{x}\rangle}\in\alg{T}$ and so beable status for $|v\rangle$ requires
$A\bullet
 P_{|v_{x}\rangle}\in\alg{I}_{|v\rangle}$, i.e. $A\bullet
 P_{|v_{x}\rangle}|v\rangle=|0\rangle$ (Thms. \ref{quasicommutative},
 \ref{defsubalg}).  Since
 $P_{|v_{x}\rangle}|v\rangle=c|v_{x}\rangle$
 (with $c\not=0$, by $3.$), using the definition of the antisymmetric
 product (Eqn. \ref{eq:antisymmetric}) yields
 \begin{equation}
 cA|v_{x}\rangle=P_{|v_{x}\rangle}(A|v\rangle)\ (=c'|v_{x}\rangle,\
 \mbox{for some}\ c')
 \end{equation}
  which shows that $A$ has
 $|v_{x}\rangle$ as an eigenvector.  Since this is true for any $x\in
 X$, $A$ lies in the definite sets of all the states
 $\{|v_{x}\rangle\}$, and therefore $A\in\alg{B}_{\{|v_{x}\rangle\}}$
 contrary to hypothesis.

 For the converse claim, suppose $\alg{B}\subseteq\alg{S}(H_{n})$
 is a maximal beable subalgebra for the state $|v\rangle\in H_{n}$.
Consider the
 subspace of $H_{n}$ given by
 \begin{equation}
 S=\{|w\rangle\in H_{n}:\ (A\bullet B)|w\rangle=|0\rangle\ \mbox{for all}\
 A,B\in\alg{B}\}
 \end{equation}
which is nontrivial since $\alg{B}$'s beable status for $|v\rangle$
requires that $|v\rangle\in S$.
We first show that $S$ is invariant under
$\alg{B}$, i.e. $|w\rangle\in S$ implies $C|w\rangle\in S$ for any
$C\in\alg{B}$.  (In fact, for this part of the argument the dimension
of the Hilbert space needn't be finite.)
To establish this, we need to show that if $|w\rangle\in
S$, then $(A\bullet B)(C|w\rangle)=|0\rangle$ for any $A,B\in\alg{B}$.
Using the fact that the $C^{\star}$ product of two
operators $X$ and $Y$ is expressible as $XY=X\circ Y-iX\bullet Y$
(see the remarks following Eqn. \ref{eq:parts}),
together with the supposition that antisymmetric products formed in $\alg{B}$
map $|w\rangle$ to zero and the definition of the
symmetric product (Eqn. \ref{eq:symmetric}), one calculates
\begin{eqnarray}
\label{eq:first}
(A\bullet B)C|w\rangle & = & ((A\bullet B)\circ C)|w\rangle
-i((A\bullet B)\bullet C)|w\rangle \\
& = & ((A\bullet B)\circ C)|w\rangle  \\
 & = & 1/2(A\bullet B)C|w\rangle+1/2C(A\bullet
 B)|w\rangle \\ \label{eq:last} & = & 1/2(A\bullet B)C|w\rangle.
  \end{eqnarray}
But Eqns. \ref{eq:first} and \ref{eq:last} are consistent only if
$(A\bullet B)(C|w\rangle)=|0\rangle$, as required.

Now, by the definition of $S$, all the operators in $\alg{B}$ commute
on $S$.  And since $S$ is invariant under $\alg{B}$, restricting the action
of any
self-adjoint operator in $\alg{B}$ to $S$ induces a self-adjoint
operator on the subspace $S$.  Since $H_{n}$---and thus $S$---is
finite-dimensional, it follows by a well-known result that the operators in
$\alg{B}$ share at least one
complete set of common eigenvectors $\{|v_{y}\rangle\}$ ($y\in Y$) \emph{on
the subspace}
$S$.  The set $\{|v_{y}\rangle\}$ clearly satisfies $1.$,
and also $2.$ since $|v\rangle$ lies in $S$.  We can also arrange
for $3.$ to be satisfied---while preserving satisfaction of $1.$ and
$2.$---by just dropping from the set $\{|v_{y}\rangle\}$
any vectors orthogonal to $|v\rangle$.  So we can conclude, then, that
there is at least one set
of vectors $\{|v_{x}\rangle\}$ ($x\in X$) satisfying $1.$--$3.$ which are
common
eigenvectors of all the beables in $\alg{B}$.  If so, then clearly
$\alg{B}\subseteq\alg{B}_{\{|v_{x}\rangle\}}$, and the hypothesis that
$\alg{B}$ is
maximal for $|v\rangle$ delivers the required conclusion
that $\alg{B}=\alg{B}_{\{|v_{x}\rangle\}}$ for some set satisfying $1.$-$3$.
(For uniqueness of this set, it is easy to see that if
$\alg{B}_{\{|v_{x}\rangle\}}=\alg{B}_{\{|v_{y}\rangle\}}$ for two sets
of vectors satisfying $1.$--$3.$, then those sets must in fact
generate the same rays in $H_{n}$.)\ \emph{QED}.\vspace{.15in}

\noindent For finite-dimensional $H$, Thm. \ref{vectormaximal} yields a
complete
picture of maximal subalgebras for any pure state on $\alg{S}$:
they simply correspond 1-1 with sets of vectors satisfying $1.$--$3.$
of the theorem, which then end up being common eigenvectors for all the elements of
the algebra.  If the set contains only a single vector, we get the
orthodox subalgebra; if the set is a basis for
the Hilbert space, we get a commutative subalgebra; and if the set falls
between these two extremes we get a subalgebra of Bub-definite
type.

For infinite-dimensional $H$, the converse part of Thm.
\ref{vectormaximal} proof
breaks down at the point where the existence of a complete set of
commuting eigenvectors on the finite-dimensional subspace $S$ is
invoked.  For if $S$
 could no longer be assumed finite-dimensional, some of the elements of
$\alg{B}$
 might then have \emph{no}
eigenvectors in $S$, much less any common ones.  However, I
conjecture that in that case there still exists a family $\{\rho_{x}\}$ of
\emph{singular} pure states on
$\alg{S}(H)$ satisfying
$1.$-$3.$ with respect to $|v\rangle$,
where by that I mean states that are not
representable by vectors in $H$, such as the Dirac states I invoked
in the last section.\footnote{Note that $1.$-$3.$ make sense
for families of states on $\alg{S}(H)$ whether or not they are singular.
For orthogonality of two algebraic states is equivalent to the norm of
their difference (as a linear
functional) being 2 (Sakai 1971, p. 31), and $2.$ is equivalent to
the assertion that any state orthogonal
to all the $\rho_{x}$'s is orthogonal to $|v\rangle$ (thanks to
Laura Reutsche for this point).}  Furthermore, I suspect that if $1.$--$3.$ are
satisfied by singular pure states $\{\rho_{x}\}$,
$\alg{B}_{\{\rho_{x}\}}$
\emph{is} a maximal beable
subalgebra for $|v\rangle$---though obviously Thm.
\ref{vectormaximal}'s maximality
proof no longer works since it relies on assuming that the pure
states satisfying $1.$--$3.$ are vector states in the ranges of
one-dimensional
projection operators.

I end this section with one last (rather suggestive) result that
illustrates the
possibility of a maximality argument without assuming the state is a
vector state.

  \begin{theorem}
 \label{maximal}
 The definite set of a singular pure state $\rho$ on
 $\alg{S}(H)$ is
 a maximal beable subalgebra for $\rho$.
 \end{theorem}
 \emph{Proof}.  I need a nontrivial result due to Kadison and
 Singer (1959, Thm. 4).
 They show that the definite set of any pure state $\rho$ on the
 $C^{\star}$-algebra $\alg{U}(H)$ of all bounded operators on a Hilbert space
 (defined as the set of all \emph{self-adjoint} operators on $H$
 on which
 $\rho$ is dispersion-free) is not properly
 contained in the definite set of any other state on $\alg{U}(H)$.
   It is not difficult to see that a state is
 pure on a Segalgebra $\alg{S}$ exactly when its extension to the
$C^{\star}$-algebra
 generated by $\alg{S}$ is pure.  So the Kadison-Singer
 maximality result also holds for the pure states on
 $\alg{S}(H)$, and in particular the singular pure states.

 Now consider the definite set $\alg{D}_{\rho}$ of a singular pure
 state $\rho$ on $\alg{S}(H)$.  To show $\alg{D}_{\rho}$ is a
 maximal beable subalgebra for $\rho$, let $\alg{Q}$ denote the subalgebra
 generated by $\alg{D}_{\rho}$ together with any element
 $A\not\in\alg{D}_{\rho}$.  Note that $I-A\not\in\alg{D}_{\rho}$,
 and either $A$ or $I-A$ must lie outside
 of $\alg{I}_{\rho}$ (otherwise both would get value $0$ in state $\rho$,
 yet their values must sum to $1$).  So there is a $B\in\alg{Q}$ such that
 $B\not\in\alg{D}_{\rho}$ and $B\not\in\alg{I}_{\rho}$.  Now suppose
 $\alg{Q}$ has beable status for $\rho$.  Then Thm.
 \ref{quasicommutative} (Part 2.) and Thm. \ref{defsubalg} dictate that
there is a
 dispersion-free state $\val{\cdot}$ on $\alg{Q}$ such that
 $\val{B}\neq0$.  Using a Hahn-Banach-type argument,
one can show that any state of a sub-Segalgebra
extends to a state on the whole algebra.\footnote{See Segal (1947, Lemma
2.2) or
Kadison and Ringrose (1983, Thm. 4.3.13).}
Therefore, $\val{\cdot}$ extends to a state on $\alg{S}(H)$ that has
a definite set properly containing $\rho$'s---contradicting the
Kadison-Singer maximality result.\ \emph{QED}.\vspace{.15in}

\section{From Quasicommutative to Commutative}

In this final section, I want to point out two ways of arguing
for the commutativity of beables.  The first way depends on rejecting
the modal idea of letting a system's beables vary from one of its
quantum states to the next, and the second way depends on
the Reeh-Schlieder theorem of algebraic relativistic
quantum field theory.

First, suppose one had reasons to demand that a quantum system possess a
fixed set of beables
for all its quantum states.  For example, one could think that the idea of a
physical magnitude having a definite value at one time but not at
another is conceptually incoherent (though modalists would dispute this).
Or one could think that the extra flexibility of
having a state-dependent set of beables is not necessary;
in particular, not needed to solve the
measurement problem, since all measurement outcomes can
be ensured simply by granting beable status to (essentially) a single
observable,
like position.\footnote{This is a view Bell seems to have held---cf. the
second quotation of Section
1.}

Now it would be unreasonable to require a fixed set of beables to satisfy
the statistics of \emph{all} states, since they may not all be
realizable in nature.  An example is provided by Dirac delta
states, which we saw a few sections back qualify as
algebraic states.  Obviously to actually prepare
such a state would require infinite energy on pain of violating the uncertainty
relation.\footnote{A more sophisticated example
of the fact that not all algebraicly defined states are necessarily
physical is found in algebraic quantum field theory on curved
spacetime.  There, the expectation value of the stress-energy
tensor is not defined for all algebraic states of the
field, but only for the so-called `Hadamard' states; see Wald (1994, Sec.
4.6).}
However, even without committing to definitive necessary and sufficient
conditions for a state to count as physical, it seems reasonable to
expect a system's physical states to make up a full
set of states
on the system's Segalgebra, as do the vector states of
$\alg{S}(H)$.\footnote{For the Hadamard states referred
to in the previous note, Thm. 2.1 of Fulling et
al. (1981)
 establishes that they
are dense in any Hilbert space representation of a globally hyperbolic
spacetime's Segalgebra of
observables, from which it follows by an
elementary argument that
the set of Hadamard states is full too.}  Assuming this, we have
the following converse to Thm. \ref{every} establishing
commutativity for nonmodal interpreters who want a state-independent
ontology of a system's properties (as in Bohm's theory).

\begin{theorem}
\label{nogo}
If a subalgebra of a Segalgebra $\alg{S}$ has beable status for every state
in a
full set of states on $\alg{S}$, then it is commutative.
\end{theorem}
\emph{Proof}.  Let $\alg{B}$ be a subalgebra of $\alg{S}$ and $\Omega$ a full
set of states on $\alg{S}$.  If $\alg{B}$ has beable status for every
state in $\Omega$, Thms. \ref{quasicommutative} (Part $3.$) and \ref{defsubalg}
dictate that $\alg{B}\bullet\alg{B}\subseteq\alg{I}_{\omega}$ for
all $\omega\in\Omega$.  But since $\Omega$ is a full set, it is easy
to see that $\bigcap_{\omega\in
\Omega}\alg{I}_{\omega}=\{0\}$, which forces
$\alg{B}$ to be commutative.\ \emph{QED}.\vspace{.15in}

Thm. \ref{nogo} also allows us to diagnose exactly what goes wrong in
Misra's (1967)
argument against hidden-variables (in fulfillment of a promise I made
a few sections back).  Without distinguishing
`observables' whose measurement outcomes are
determined by hidden-variables from those which correspond to true
beables of the system, Misra demands that the outcome of
any measurement be determined by a dispersion-free state on the
algebra of all \emph{observables} of a system (1967, p.
856), which we've seen is only reasonable if all `observables' are beables
of the system.
Misra also assumes that the hidden-variables must be adequate for
recovering (after averaging) the expectation values of at least the
physical states of a
system, which he too assumes will be a full set.  That this is a lethal
combination
of assumptions should be clear.  If all
`observables' are treated as beables, and they are forced to satisfy the
statistics of a full set of states of the system, then Thm. \ref{nogo}
dictates that all the
\emph{observables} of the system have to commute---which is absurd!  Far
from delivering a no-go
theorem, this is simply one more confirmation of Bell's point that not all
`observables'
can have beable status.

A second way to argue that beables must be commutative arises out of the
algebraic approach to relativistic
quantum field theory.  In that approach,
one associates with each bounded open region $O$ in Minkowski
spacetime $M$ a $C^{\star}$-algebra $\alg{U}(O)$ whose Segalgebra represents
all observables measurable in region $O$.  In the `concrete' approach,
the algebras $\{\alg{U}(O)\}_{O\subseteq M}$ are taken to be von Neumann
algebras of
operators
acting on some common Hilbert space consisting of states of the
entire field on $M$.  If the collection of algebras
$\{\alg{U}(O)\}_{O\subseteq M}$ satisfies a number of very general
assumptions involving locality, covariance, etc. (the details of which need
not detain us here),
then it becomes possible to
prove the Reeh-Schlieder theorem whose main consequence is that
every state vector of the field with bounded energy is a \emph{separating
vector} for all the local algebras $\{\alg{U}(O)\}_{O\subseteq M}$.
This means that no nonzero operator $A$ in any local algebra $\alg{U}(O)$ can
annihilate such a state
vector.\footnote{Haag (1992, Sec. II.5.3); see also Redhead (1995b) for an
elementary
discussion.}

Now consider this result in the context of my analysis of
beable subalgebras of the observables of a
system.  Here the role of the system is played by the quantum field
in some bounded open region $O$ of spacetime, and the question is
which of the observables in $\alg{U}(O)$'s Segalgebra can be granted
beable status.  As we've seen, a subalgebra $\alg{B}(O)$ will have beable
status for a state $\omega$ of the field if and only if $\alg{B}(O)$
is $\alg{I}_{\omega|_{\alg{B}(O)}}$-quasicommutative.  But if $\omega$
corresponds to
a state vector $|v\rangle$ with bounded energy, then since that
vector is separating for $\alg{U}(O)$, we have
\begin{equation}
A\in\alg{I}_{\omega|_{\alg{B}(O)}}\Leftrightarrow\langle
v|A^{2}|v\rangle =0\Leftrightarrow\langle Av|Av\rangle
=0\Leftrightarrow A|v\rangle =|0\rangle\Leftrightarrow A=0.
\end{equation}
It follows that:
\begin{theorem}
\label{fields}
 Subalgebras of local beables selected from the Segalgebras
of local observables in relativistic quantum field theory must be
commutative in any state of the field with bounded energy.
\end{theorem}

Notice that the orthodox approach to value definiteness
 reduces to absurdity in this context: since
$\omega$ is dispersion-free on $A\in\alg{U}(O)$ exactly when
\begin{equation}
A-\omega(A)I\in\alg{I}_{\omega|_{\alg{B}(O)}}=\{0\},
\end{equation}
taking $\alg{B}(O)$ to be the definite set
of $\omega|_{\alg{B}(O)}$ yields only multiples of the identity
operator as beables!  Of course,  there is
still plenty of room left for the realist to propose beable status for other
more satisfactory sets
of commuting
local observables.\vspace{.15in}

\noindent \emph{Acknowledgements}---I would like to
thank John L. Bell and Alex Wilce for occasional guidance on matters
mathematical.\vspace{.15in}

 \begin{center}
 \textbf{References}
 \end{center}

\footnotesize{\noindent Bell, J. L. and Clifton, R.  (1995)  `QuasiBoolean
Algebras
and Simultaneously Definite Properties in Quantum Mechanics',
\emph{International Journal of Theoretical Physics} \textbf{34},
2409-21.\vspace{.1in}

\noindent Bell, J. S.  (1987) \emph{Speakable and Unspeakable in Quantum
Mechanics}  (Cambridge:
Cambridge University Press).\vspace{.1in}

\noindent Bohm, D. (1952) `A Suggested Interpretation of the
Quantum Theory in Terms of ``Hidden Variables'', Part II',  \emph{Physical
Review} \textbf{85}, 180-93.\vspace{.1in}

\noindent Bratteli, O. and Robinson, D. W. (1987)  \emph{Operator Algebras
and Quantum Statistical Mechanics, Vol I.}  (Berlin: Springer-Verlag,
2nd edition).\vspace{.1in}

\noindent Bub, J. (1997) \emph{Interpreting the Quantum World}
(Cambridge: Cambridge University Press).\vspace{.1in}

\noindent Bub, J. and Clifton, R. (1996) `A Uniqueness Theorem for
``No Collapse'' Interpretations of Quantum Mechanics',  \emph{Studies in
History and Philosophy of Modern Physics} \textbf{27}, 181-219.\vspace{.1in}

\noindent Clifton, R.  (1995)  `Independently Motivating the
Kochen-Dieks Modal Interpretation of Quantum Mechanics',  \emph{British
Journal for Philosophy of Science}  \textbf{46}, 33-57.\vspace{.1in}

\noindent Clifton, R. (1996) `The Properties of Modal
Interpretations of Quantum Mechanics',  \emph{The British Journal for
Philosophy of Science} \textbf{47}, 371-398.\vspace{.1in}

\noindent D\"{u}rr, D., Goldstein, S. and Zanghi, N. (1996)
`Bohmian Mechanics as the Foundation of Quantum Mechanics',  in J. T.
Cushing et al. (eds.), \emph{Bohmian Mechanics and Quantum Theory: An
Appraisal}  (Dordrecht: Kluwer),  pp. 21-44.\vspace{.1in}

\noindent Fulling, S. A., Narcowic, F. J. and Wald, R. M. (1981)
`Singularity Structure of the Two-Point Function in Quantum Field
Theory in Curved Spacetime, II', \emph{Annals of Physics} \textbf{136},
243-72.\vspace{.1in}

\noindent Geroch, R. (1985) \emph{Mathematical Physics}  (Chicago:
University of Chicago Press).\vspace{.1in}

\noindent Haag, R.  (1992) \emph{Local Quantum Physics: Fields,
Particles and Algebras} (Berlin: Springer-Verlag).\vspace{.1in}

\noindent Heywood, P. and Redhead, M. L. G. (1983) `Nonlocality and
the Kochen-Specker Paradox',  \emph{Foundations of Physics} \textbf{13},
481-99.\vspace{.1in}

\noindent Kadison, R. V. (1975)  `Operator Algebras',  in J. H.
Williamson (ed.), \emph{Algebras in Analysis}  (London: Academic
Press), pp. 101-17.\vspace{.1in}

\noindent Kadison, R. V. and Ringrose, J. R. (1983)  \emph{Fundamentals
of the Theory of Operator Algebras.  Vol. 1}  (London: Academic
Press).\vspace{.1in}

\noindent Kadison, R. V. and Singer, I. M. (1959)  `Extensions of
Pure States',  \emph{American Journal of Mathematics} \textbf{81},
383-400.\vspace{.1in}

\noindent Landsman, N. P. (forthcoming) \emph{Tractatus
Classico-Quantum Mechanicus}  (Berlin: Springer-Verlag).\vspace{.1in}

\noindent Misra, B. (1967)  `When Can Hidden Variables be Excluded
in Quantum Mechanics?', \emph{Il Nuovo Cimento} \textbf{47A},
841-59.\vspace{.1in}

\noindent Pagonis, C. and Clifton, R. (1995)  `Unremarkable
Contextualism: Dispositions in the Bohm Theory',  \emph{Foundations of
Physics} \textbf{25}, 281-296.\vspace{.1in}

\noindent Plymen, R. J.  (1968) `Dispersion-Free Normal
States', \emph{Il Nuovo Cimento} \textbf{54}, 862-70.\vspace{.1in}

\noindent Redhead, M. L. G. (1987) \emph{Incompleteness, Nonlocality and
Realism}  (Oxford: Clarendon Press).\vspace{.1in}

\noindent Redhead, M. L. G. (1995a)  \emph{From Physics to Metaphysics}
(Cambridge: Cambridge University Press).\vspace{.1in}

\noindent Redhead, M. L. G. (1995b)  `More Ado About Nothing',
\emph{Foundations of Physics} \textbf{25}, 123-37.\vspace{.1in}

\noindent Rudin, W. (1974)  \emph{Real and Complex Analysis} (New York:
McGraw-Hill, 2nd ed).\vspace{.1in}

\noindent Sakai, S. (1971) \emph{$C^{\star}$-algebras and
$W^{\star}$-algebras} (Berlin: Springer).\vspace{.1in}

\noindent Segal, I.  (1947) `Postulates for General Quantum
Mechanics', \emph{Annals of Mathematics} \textbf{4}, 930-948.\vspace{.1in}

\noindent Simmons, G. F.  (1963)  \emph{Introduction to Topology and Modern
Analysis}  (New York: MacGraw-Hill).\vspace{.1in}

\noindent Van Fraassen, B. C. (1973)  `Semantic Analysis of Quantum
Logic',  in C. A. Hooker (ed.), \emph{Contemporary Research in the
Foundations and Philosophy of Quantum Theory}  (Dordrecht: Reidel),
pp. 80-113.\vspace{.1in}

\noindent Wald, R. M. (1994)  \emph{Quantum Field Theory in Curved
Spacetime and Black Hole Thermodynamics}  (Chicago:
University of Chicago Press).\vspace{.1in}

\noindent Zimba, J. and Clifton, R. (forthcoming) `Valuations on
Functionally Closed Sets of Quantum Mechanical Observables and Von
Neumann's ``No-Hidden-Variables'' Theorem',  in P. Vermaas and D. Dieks
(eds.), \emph{The Modal Interpretation of Quantum Mechanics}}
(Dordrecht: Kluwer).

\end{document}